\documentclass[10pt,conference]{IEEEtran}

\usepackage{amsmath,amssymb,amsfonts}
\usepackage{nicefrac}
\usepackage{braket}

\usepackage{graphicx}
\usepackage{textcomp}
\usepackage{booktabs}
\usepackage{multirow}
\usepackage{makecell}
\usepackage{adjustbox}
\usepackage{microtype}
\usepackage{setspace}
\usepackage{enumitem}

\usepackage{xcolor}
\usepackage[most]{tcolorbox}
\tcbuselibrary{listings,skins,breakable}

\usepackage{listings}

\usepackage{algorithm}
\usepackage{algpseudocode}

\usepackage{tikz}
\usetikzlibrary{arrows.meta,positioning,fit,calc}

\usepackage{orcidlink}
\usepackage{comment}
\usepackage{url}

\usepackage{hyperref}
\usepackage{cleveref}

\newtcolorbox[auto counter, number within=subsection]{mybox}[2][]{%
  enhanced,
  colback=black!3!white,
  colframe=black!90!black,
  breakable,
  left=0.5pt,
  right=0.5pt,
  top=0.5pt,
  bottom=0.5pt,
  boxsep=0.5pt,
  boxrule=0.5pt,
  title=Box~\thetcbcounter\ -- #2,
  label type=mybox,
  #1
}

\crefname{mybox}{Box}{Boxes}
\Crefname{mybox}{Box}{Boxes}

\def\BibTeX{{\rm B\kern-.05em{\sc i\kern-.025em b}\kern-.08em
    T\kern-.1667em\lower.7ex\hbox{E}\kern-.125emX}}

\begin{document}

\title{Q-LEAK: Quantum-Based LEAKage Verification for\\ Side-Channel Countermeasures

}

\author{\IEEEauthorblockN{Walid El Maouaki\IEEEauthorrefmark{1}\IEEEauthorrefmark{2}\orcidlink{0009-0004-2339-5401}, Alberto Marchisio\IEEEauthorrefmark{1}\IEEEauthorrefmark{2}\orcidlink{0000-0002-0689-4776},
Muhammad Shafique\IEEEauthorrefmark{1}\IEEEauthorrefmark{2}\orcidlink{0000-0002-2607-8135}}

\IEEEauthorblockA{\IEEEauthorrefmark{1} \normalsize eBrain Lab, Division of Engineering, New York University Abu Dhabi, PO Box 129188, Abu Dhabi, UAE\\}
\IEEEauthorblockA{\IEEEauthorrefmark{2} \normalsize Center for Cyber Security, NYUAD Research
Institute, New York University Abu Dhabi, UAE\\
Emails: walid.el.maouaki@nyu.edu, alberto.marchisio@nyu.edu, muhammad.shafique@nyu.edu}
}

\maketitle


\begin{abstract}

Formal verification of power side-channel leakage and its countermeasures in cryptographic algorithms is challenging, as SAT-based methods fail to scale on XOR-heavy, time-unrolled cryptographic circuits with realistic leakage models. We construct compact Conjunctive Normal Form (CNF) cases modeling one-bit leakage under two-trace conditions, linking key dependence and state evolution. Classical solvers quickly reach complexity limits, so we propose \textit{Q-LEAK}, a quantum-based verification approach using Grover’s algorithm, compiling each CNF into an oracle and applying amplitude amplification to search in $O(\sqrt{N})$ oracle calls, with oracles that encode the two-trace leakage predicate and the CNF constraints. Benchmarking against classical SAT shows both potential gains and practical resource limits. In noiseless tests on $5-7$ variable benchmarks, \textit{Q-LEAK} consistently recovered a satisfying assignment within $1-4$ tries, 
with marked bitstrings amplified clearly above the background distribution, exceeding 20\% probability. The evaluation of \textit{Q-LEAK} on real quantum hardware revealed at least one classically verified SAT assignment, despite the presence of noise. These results point to a potential path toward quantum-assisted verification of side-channel protections.
\end{abstract}



\section{Introduction}











\begin{figure}[t!]
    \centering
    \includegraphics[width=0.9\linewidth]{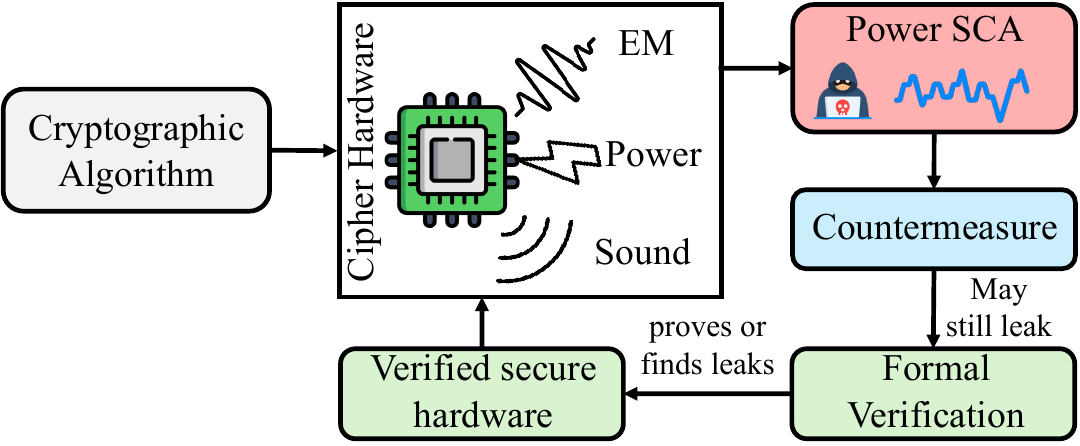}
    \caption{Cryptographic algorithm is implemented as a cipher in hardware, exposed to power side-channel attacks. Masking countermeasures are applied to protect intermediate values, and formal verification ensures that the final implementation remains leakage-free after synthesis.}
    \label{Motivation}
\end{figure}

Power side-channel leakage constitutes a persistent threat to the security of hardware-implemented cryptographic primitives~\cite{agrawal2002side, lou2021survey}. Despite the incorporation of masking and related countermeasures by hardware designers~\cite{spreitzer2017systematic, lyu2018survey}, the final synthesized circuit may still exhibit information leakage through unintended switching activity and transient glitches. Testing methodologies alone cannot exhaustively cover the exponentially large input space. Therefore, formal verification methods are required to establish \textit{leak-free} properties under an explicit leakage model. Fig.~\ref{Motivation} illustrates this verification pipeline. A cryptographic algorithm becomes a hardware cipher subject to power-monitoring adversaries; countermeasures are integrated into the design, and a formal verification step must either identify a leakage vulnerability or certify the security of the design.

\subsection{Research Problem and Challenges}
State-of-the-art formal verification methods encode the cipher, the countermeasure, and a two-trace leakage relation as a Boolean satisfiability (SAT) or satisfiability-modulo-theories (SMT) problem in conjunctive normal form (CNF), then search for assignments that violate the desired security property~\cite{coron2018formal, eldib2014formal}. These encodings are often XOR-heavy, require time-unrolling across $T$ cycles, and include key bits, data, randomness, and internal wires. The resulting assignment space scales as $N=2^n$, where $n$ is the number of Boolean variables. Modern CDCL and SMT engines exploit propagation, learning, and structure, so they should not be viewed as literal exhaustive enumerators on every instance; nevertheless, their worst-case behavior remains exponential. We therefore use the oracle-query model only to isolate the search component that Grover's algorithm accelerates, while using Kissat as the classical correctness baseline in the experiments.


\subsection{Motivation: Complexity Scaling}

\begin{figure}[t!]
    \centering
    \includegraphics[width=\linewidth]{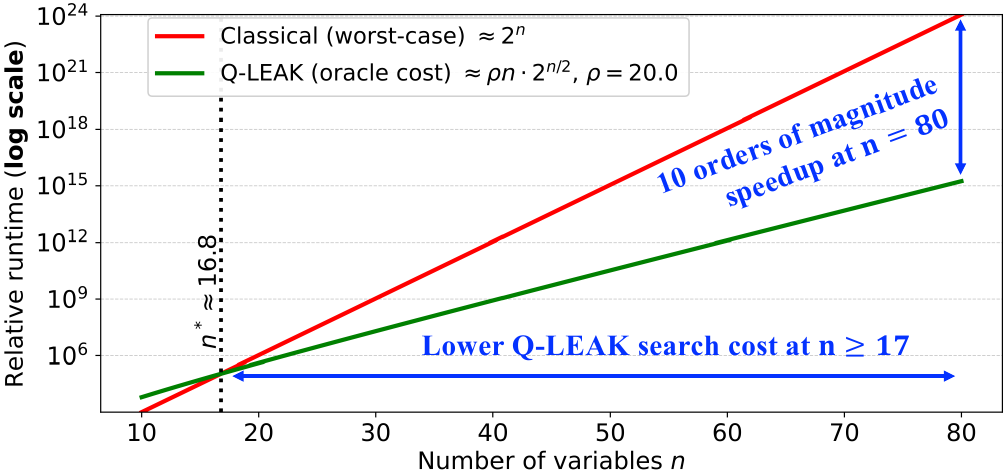}
    \caption{Asymptotic scaling of search cost versus the number of variables $n$ on a log scale: classical worst-case scaling $\sim 2^n$ (red) versus Q-LEAK with density-aware oracle cost $\sim \rho n \cdot 2^{n/2}$ (green), where $\rho=m/n$ is the clause density. For $\rho=20$, the crossover occurs near $n^\star \approx 16.8$, with Q-LEAK showing a growing relative benefit for larger $n$.}
    \label{Complexity}
    \vspace{-6pt}
\end{figure}

\begin{table}[t!]
\centering
\caption{Asymptotic comparison (worst case) for CNF-encoded side-channel verification with $n$ variables and $m$ clauses.}
\label{tab:classical-vs-grover-oracle}
\resizebox{0.95\columnwidth}{!}{%
\begin{tabular}{lllll}
\toprule
Algorithm & Queries / iters & Per-query/iter cost & Total cost \\
\midrule
\makecell{Exhaustive\\ enumeration~\cite{eldib2014formal}} & $\Theta(2^n)$ worst case & \makecell{predicate eval} & $\Theta(2^n)$ \\
\hline
\makecell{CDCL SAT\\ solver~\cite{een2005minisat, leventi2021cryptominisat}}  & $\Theta(2^n)$ worst case & \makecell{predicate / BCP /\\ learning} & $\Theta(2^n)$ \\
\hline
\makecell{SMT-based\\ verifier~\cite{eldib2014smt}}  & $\Theta(2^n)$ worst case & \makecell{predicate / BCP /\\ learning} & $\Theta(2^n)$ \\
\hline
\makecell{(\textbf{ours}) Q-LEAK\\ (unknown $K$)}   & $\Theta(\sqrt{2^n/K})$ iters
                       & \makecell{Oracle $\approx \Theta(m)$;\\ Diffuser $=\Theta(n)$}
                       & $\Theta\!\big(m\sqrt{2^n/K}\big)$ \\
\bottomrule
\end{tabular}%
}
\end{table}

Classical formal verification methodologies for power-leakage security exhibit poor scalability, may require search over up to $2^n$ assignments in the worst case. for $n$ key/state variables in the worst-case scenario. Fig.~\ref{Complexity} and Table~\ref{tab:classical-vs-grover-oracle} present a comparative complexity analysis that identifies a critical computational gap: Grover's quantum search algorithm operates in $\Theta(\sqrt{N})$ queries to the verification predicate $f$, where $N=2^n$, establishing the foundation for accelerated leakage verification at larger problem sizes.

Within the \textit{Q-LEAK} framework, the two-trace bit-level leakage CNF is encoded into a reversible Grover oracle whose cost scales with the number of CNF clauses. We express this clause growth using the density $\rho=\frac{m}{n}$, where $m$ is the number of clauses and $n$ is the number of variables; therefore, $m=\rho n$. Under this density-based formulation, the oracle-aware Q-LEAK runtime is modeled as $T_{\text{Q-LEAK}}(n,\rho)\approx \rho n \cdot 2^{n/2}$, where the factor $2^{n/2}$ comes from Grover's quadratic reduction of the search space and the multiplicative term $\rho n$ accounts for the reversible evaluation cost of the CNF clauses. For $k$ satisfying leakage witnesses, the corresponding expression becomes $$T_{\text{Q-LEAK}}(n,\rho,k)\approx \frac{\rho n \cdot 2^{n/2}}{\sqrt{k}}.$$

The reversible oracle construction uses approximately one ancilla qubit per CNF clause, together with the $n$ variable qubits and one output/phase qubit. Hence, the density-dependent qubit requirement is $$n+m+1=n+\rho n+1=(1+\rho)n+1.$$ This shows that the density $\rho$ directly controls both the oracle-evaluation cost and the number of auxiliary qubits required by the Q-LEAK circuit.

The crossover point between the classical worst-case scaling and the Q-LEAK oracle-aware scaling is obtained from $2^n=\rho n 2^{n/2}$, equivalently $2^{n^\star/2}=\rho n^\star$. Thus, for a given density $\rho$, Q-LEAK dominates once $n>n^\star$.

In Fig.~\ref{Complexity}, the density is set to $\rho=20$, corresponding to $m=20n$ clauses. The red curve represents the classical worst-case runtime, $$T_{\text{classical}}(n)\approx 2^n,$$ whereas the green curve represents the Q-LEAK oracle-aware runtime, $$T_{\text{Q-LEAK}}(n)\approx \rho n 2^{n/2}.$$ The vertical dotted line marks the density-based crossover at approximately $n^\star \approx 16.8$. Accordingly, the plot highlights that Q-LEAK enters its advantage regime for problem sizes beyond roughly $n>15$.

At $n=80$ variables with density $\rho=20$, the corresponding clause count is $m=\rho n = 20 \times 80 = 1600$.
At this scale, Fig.~\ref{Complexity} shows a separation of roughly ten orders of magnitude between the classical worst-case curve and the Q-LEAK oracle-cost curve. This illustrates the central scaling advantage: although the Q-LEAK oracle cost grows linearly with the clause density through the factor $\rho n$, the dominant search term is reduced from $2^n$ to $2^{n/2}$, producing an exponentially widening gap as $n$ increases.

This density-based complexity profile motivates the Q-LEAK verification pipeline: a CNF-to-oracle construction coupled with Grover search that preserves the quadratic $\sqrt{N}$ speedup while explicitly accounting for clause-density overhead. The resulting model demonstrates that, for fixed clause density $\rho$, Q-LEAK achieves a decisive asymptotic advantage once the number of variables exceeds the density-dependent crossover threshold $n^\star$.

\subsection{Novel Contributions and System Overview}

We introduce \textit{Q-LEAK}, a quantum-based formal verification framework for side-channel countermeasure analysis. The protected cryptographic implementation and associated leakage constraints are compiled to a reversible phase oracle $U_F$, upon which amplitude amplification is applied to the variable qubits encoding the search space. Specifically, we employ the Boyer-Brassard-Høyer-Tapp (BBHT) variant of Grover's algorithm when the cardinality of satisfying assignments $K$ is unknown. The quantum search mechanism operates by exploring all candidate variable assignments in superposition; the oracle applies a phase inversion to the assignments that correspond to security violations; and interference amplifies the probability amplitudes of solution states while destructively interfering with non-solution states. The contribution lies in the leakage-aware CNF-to-oracle flow, the end-to-end simulation and hardware demonstration, and the explicit resource analysis, building on the standard Grover search framework.

Our experimental results demonstrate that this approach produces sharply peaked probability distributions concentrated at leakage-inducing assignments for satisfiable cases, and flat spectral distributions for unsatisfiable cases, thereby aligning with the verification objectives of either establishing security certifications or identifying leakage vulnerabilities. By pairing standard CNF encoding methodologies with a Grover-based quantum search engine, our \textit{Q-LEAK} framework directly addresses the computational bottleneck identified in Table~\ref{tab:classical-vs-grover-oracle}, offering a motivated route toward end-to-end formal verification of power side-channel countermeasures.

In summary, the primary novel contributions of this work are as follows (see Fig.~\ref{fig:contribution}):

\begin{figure}[t!]
    \centering
    \includegraphics[width=1\linewidth]{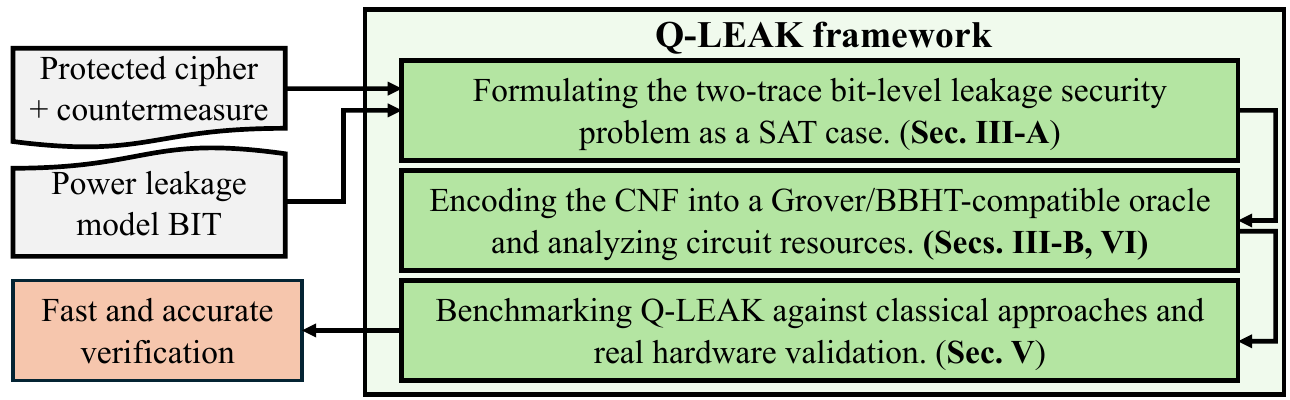}
    \caption{Overview of the main contributions of Q-LEAK. Green boxes highlight the proposed components.}
    \label{fig:contribution}
\end{figure}

\begin{itemize}
\item We present \textit{Q-LEAK}, an end-to-end leakage-verification flow that connects two-trace CNF side-channel encodings with Grover/BBHT-style quantum search. The novelty lies in integrating leakage-aware formal verification with a concrete reversible oracle construction and validation pipeline.

\item We formulate the two-trace bit-level power leakage security problem as a Boolean satisfiability case over secret keys, internal states, and time-unrolled circuit logic. This formulation enables CNF encoding within a quantum circuit compatible with Grover's algorithm.

\item We provide a rigorous theoretical analysis indicating a conditional oracle-query advantage. for side-channel verification, establishing a cost model that shows classical solvers require \(\Theta(2^n)\) time complexity while our quantum-based \textit{Q-LEAK} approach achieves \(\Theta(m \cdot 2^{n/2})\) complexity, with the crossover point occurring at residual problem sizes predicted by the selected density model.

\item  We evaluate \textit{Q-LEAK} on small, reproducible SAT/UNSAT benchmark instances and compare every returned witness against the Kissat classical SAT solver. These experiments validate correctness rather than claiming that current quantum runs outperform classical solvers on the tested sizes.

\item We evaluate \textit{Q-LEAK} under noisy intermediate-scale quantum devices, where our framework recovered at least one valid SAT assignment despite hardware noise, although spurious peaks and lower contrast were observed, demonstrating robustness and motivating future hybrid leak/no-leak screening on residual instances where classical SAT becomes difficult.

\item 
We identify the main scalability blockers (oracle ancillas, multi-controlled-gate depth, BBHT retries, and noise) and frame larger cryptographic benchmarks, HW/HD leakage, multi-cycle unrolling, and measurement-calibrated leakage models as future work.







\end{itemize}

\section{Background and Related Work} 

\begin{figure*}
    \centering
    \includegraphics[width=\linewidth]{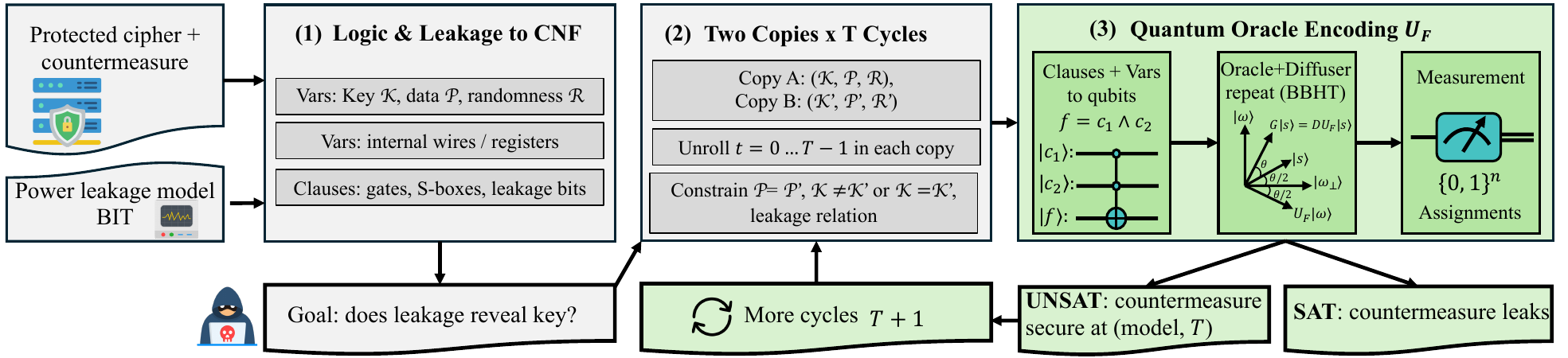}
    \caption{Q-LEAK framework. Compile the protected cipher and BIT model into a two-trace, $T$-cycle CNF, map it to an oracle $U_F$; then run Grover+BBHT over the variable qubits. Peaks indicate SAT (leak), near-uniform spectra indicate UNSAT (secure at the chosen model and $T$).
}
    \label{fig:Methodology}
\end{figure*}

While modern encryption schemes in cryptography, such as AES, DES, and RSA, are designed to resist conventional cryptanalytic attacks under standard assumptions, their hardware implementations suffer from physical leakage vulnerabilities that compromise these guarantees~\cite{mai2011side}. Devices executing secure algorithms inadvertently leak sensitive information through observable physical characteristics (e.g., power consumption and electromagnetic (EM) emanations), which can be exploited through side-channel attacks~\cite{kocher1999differential}. These attacks correlate physical measurements with internal cryptographic states using leakage models such as bit-level models~\cite{messerges2002examining}, Hamming weight models that count the number of bits set to one, and Hamming distance models that measure the number of bit transitions~\cite{brier2004correlation,mangard2007power}. To mitigate these vulnerabilities, designers employ countermeasures such as masking, hiding, and noise injection~\cite{nikova2006threshold,chari1999towards}. Their effectiveness is assessed through formal verification, which translates masked hardware into CNF or Boolean formulas for leakage analysis~\cite{eldib2014formal,barthe2015verified}. Classical SAT solvers are then applied to mathematically prove that no exploitable correlation exists between observable side channels and secret keys~\cite{bloem2018formal,hadzic2021cocoalma}. However, this verification process remains computationally demanding, as SAT solvers exhibit exponential behavior when analyzing complex masking schemes, creating practical limits on the order of masking and circuit size that can be verified within reasonable time constraints~\cite{schreiber2024trusted}.

Grover's algorithm provides a quadratic reduction in black-box query complexity for unstructured search: one of $K$ marked elements in a domain of size $N$ can be found with $O(\sqrt{N/K})$ oracle calls, and this query bound is asymptotically optimal~\cite{grover1996fast,bennett1997strengths}. \textit{This complexity measures oracle calls rather than the wall-clock cost of constructing and executing a structured SAT oracle.}

Existing quantum SAT work addresses complementary layers of the problem. One line of work optimizes generic CNF-oracle synthesis by trading ancillary qubits against circuit size or depth~\cite{yang2023efficient}; its focus is oracle implementation rather than an application-level verification property or the outer search schedule. Another constructs comparator-based Grover oracles for small integer-domain constraint formulas and reports simulator and NISQ experiments using a predetermined, small number of Grover iterations~\cite{vinod2021finding}. A further approach parallelizes generic Boolean clause evaluation using entanglement and extends the construction across distributed nodes through teleportation; its repetition formula depends on the number of solutions, with quantum counting proposed when that number is unknown~\cite{lin2023parallel}. These works primarily address general constraint satisfaction and search problems, often assuming either known or bounded solution counts, and do not explicitly consider problem classes characterized by unknown assignment cardinalities or specialized relational constraints between multiple evaluations.

Q-LEAK differs at the integration layer by converting a leakage-aware two-trace CNF into a compute–phase–uncompute oracle, interpreting satisfying assignments as leakage witnesses, validating measured candidates classically, and using BBHT to handle unknown (K) without fixing a single Grover iteration count~\cite{boyer1998tight}. Its oracle costs can be reduced further by incorporating the ancilla and depth optimizations~\cite{yang2023efficient} and the parallel clause-evaluation techniques~\cite{lin2023parallel}.
This approach enables (a) the treatment of SAT (leak-present) vs. UNSAT (leak-absent) cases within a single verification framework, (b) the use of constructive vs. destructive quantum interference patterns in measurement histograms as an interpretable signal, and (c) a systematic study of asymptotic and practical scaling behavior while making oracle-construction costs explicit, thereby bridging formal leakage reasoning with quantum-accelerated search.

Power Side-Channel Attacks (SCAs) exploit the empirical observation that a cipher’s power consumption depends on the internal data being processed during cryptographic operations. By recording and analyzing power traces, an adversary can attempt to recover the secret key. To reason about such attacks in a principled manner, leakage models are employed to approximate the functional relationship between power consumption and internal states. In the bit-level (BIT) model, the attacker observes a single internal bit (for example, bit 3 of a register). In the Hamming-Weight (HW) model, power consumption is modeled as being proportional to the number of ‘1’ bits in the state. In the Hamming-Distance (HD) model, power is modeled as tracking the number of bit transitions between two successive states, capturing the dynamic switching activity. Simple illustrative examples show that different keys can induce distinct bit patterns, Hamming weights, or Hamming distances. Hence, an attacker can distinguish keys by observing these power traces and exploiting statistical correlations with the underlying secret data.










\section{Q-LEAK Framework}
\label{sec:methodology}

Fig.~\ref{fig:Methodology} shows the Q-LEAK flow. Starting from a protected cipher and the given BIT power-leakage model, (1) we encode logic and leakage into a CNF over key, data, randomness, and
internal wires. (2) The design is duplicated into two traces and unrolled for $T$ cycles; we tie plaintexts ($P = P'$), set key constraints, and impose the chosen leakage relation to ask: \emph{Does leakage reveal the key?} (3) This CNF is compiled into a 
reversible oracle $U_F$ and queried with the Grover/BBHT loop, then measurements
return candidate assignments. Persistent high-probability bitstrings are candidate SAT witnesses and are accepted only after classical CNF evaluation. In contrast, a near-uniform distribution is consistent with UNSAT and is treated as statistical evidence only after repeated trials and classical checks, meaning the model and the unrolling cycle $T$ do not favor any assignment; increase $T$ and repeat steps (2) and (3) until SAT is found or the resource budget is exhausted.

\subsection{Problem Formulation}

As shown in Fig.~\ref{fig:Methodology}, the formal verification of side-channel countermeasures can be formulated as deciding the satisfiability (SAT) or unsatisfiability (UNSAT) of CNF encodings that capture the cipher functionality, the associated randomness, and leakage models such as BIT, HW, or HD. A satisfiable case corresponds to the existence of a leakage-dependent behavior, which indicates vulnerability, whereas an unsatisfiable case establishes security under the chosen leakage model and unrolling round/depth. As the number of Boolean variables and the degree of temporal unrolling increase, the assignment space grows exponentially; the CNF size grows with the circuit size, leakage model, and unrolling depth, making it hard for classical CDCL and SMT solvers.

Interpreting verification as an unstructured search over all possible assignments, a predicate $F(x)$ evaluates whether an assignment violates the leakage condition. Classical brute-force query algorithms require $\Theta(2^n)$ evaluations of $F$ in the worst case, while Grover's quantum search constructs a unitary oracle $U_F$ that flags violating assignments and locates them using only $\Theta(\sqrt{2^n})$ queries (or more generally $\Theta(\sqrt{2^n / K})$ when there are $K$ satisfying assignments and $\ge 1$), in accordance with known lower bounds for unstructured quantum search.
Consequently, this provides a formal quadratic advantage for counterexample search and UNSAT detection. However, practical gains depend on oracle-construction cost, error-correction overhead, and the structural properties of the underlying CNF encodings.


Within the SAT-based verification framework, the leakage models are instantiated as CNF formulas that compare two execution traces, denoted A and B, of the same cipher operating under potentially different keys. The model (BIT, HW, or HD) defines what is observed at each round (a bit, a population count, or a transition count). These observations are then encoded into Boolean variables and constraints. A configuration parameter, referred to as the property mode, determines how the leakage observations from the two traces are related. The ``\textit{notequal}'' mode asks whether there exist two keys whose leakage differs, while the ``\textit{equal}'' mode asks whether there exist two keys whose leakage is identical. Executing a SAT solver on the CNF case for round $t$ yields the security verdict under the selected model. In ``\textit{notequal}'' mode, a SAT result indicates the existence of at least one key pair with different leakage at that round, implying key-dependent leakage and a failure of the countermeasure under that model. An UNSAT result implies that no such key pair exists, leakage is key-independent at that round, and the countermeasure holds for that model. In the ``\textit{equal}'' mode, a SAT result indicates the existence of at least one key pair that produces the same leakage, whereas a UNSAT result implies that all key pairs yield distinguishable leakage patterns.

\subsection{Quantum Circuit Construction}

Let $F(x) = \bigwedge_{j=1}^{m} C_j$ be a CNF over $n$ Boolean variables
$x \in \{0,1\}^n$, where each clause
$C_j = \bigvee_{i=1}^{p_j} \ell_{ji}$ and each literal satisfies $\ell_{ji}\in\{x_p,\neg x_p\}$.
Thus $F(x)=1$ exactly when all clauses $C_j$ are satisfied.
We implement a compute-phase-uncompute oracle $U_F$ that flips the sign if and only if
\mbox{$F(x)=1$}:
\begin{equation}
\resizebox{0.92\columnwidth}{!}{$
U_F|x\rangle|0^A\rangle = (-1)^{F(x)}|x\rangle|0^A\rangle,
\qquad
U_F = I - 2\sum_{x:F(x)=1} |x\rangle\langle x|\otimes I_A .
$}
\label{eq:UF}
\end{equation}

This is the object returned by \textsc{BuildOracle} (Alg.~1, L19-27).

\begin{figure}[t!]
    \centering
    \includegraphics[width=0.8\linewidth]{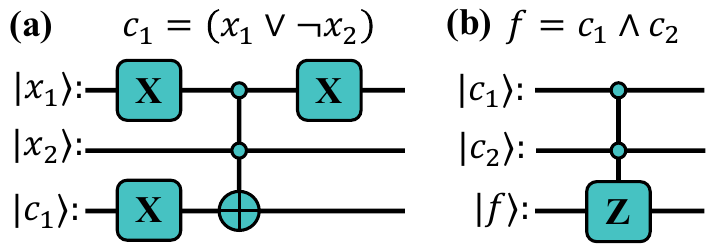}
    \caption{
    Circuit sketch for the SAT-oracle building blocks. \textbf{(a)} A clause-satisfaction bit $c_{1}$ for $C_{1}=(x_{1}\lor\neg x_{2})$ is computed reversibly by detecting the all-literals-false assignment and inverting the result. Therefore, $c_{1}=1$ exactly when clause $C_{1}$ is satisfied. \textbf{(b)} The clause-satisfaction bits $c_{1}$ and $c_{2}$ are combined using a multi-controlled operation to form the CNF flag $f=c_{1}\land c_{2}$. Thus, $f=1$ exactly when all clauses are satisfied. The ancillas are subsequently uncomputed.}
    \label{QCircuit}
\end{figure}


As shown in Fig.~\ref{QCircuit}, we allocate ancillas $c_1,\ldots,c_m$ to store the clause-satisfaction bits and an additional ancilla $f$ to store the CNF flag (Alg.~\ref{alg:grover-bbht-sca}, L21--22). For each clause $C_j$, Q-LEAK reversibly computes $c_j=C_j$, so that $c_j=1$ if and only if at least one literal in $C_j$ is true (Alg.~\ref{alg:grover-bbht-sca}, L23--24). One implementation first detects the unique clause-violating assignment in which every literal is false. For a positive literal $x_p$, surrounding $X$ gates make the corresponding control active when $x_p=0$, whereas a negated literal $\neg x_p$ is used directly as a control and is therefore active when $x_p=1$. Inverting the detected violation value then produces the clause-satisfaction bit $c_j$.

The CNF flag is subsequently computed as
$$
f \leftarrow \bigwedge_{j=1}^{m} c_j.
$$
Therefore, $f=1$ exactly when every clause is satisfied, which is equivalent to $F(x)=1$ (Alg.~\ref{alg:grover-bbht-sca}, L25). A phase flip controlled on $f$ then realizes
$$
\lvert x\rangle \mapsto (-1)^{F(x)}\lvert x\rangle
$$
(Alg.~\ref{alg:grover-bbht-sca}, L26). Finally, $f$ and all clause ancillas $c_j$ are uncomputed and returned to $\lvert 0\rangle$ (Alg.~\ref{alg:grover-bbht-sca}, L27), completing the implementation of $U_F$.
The search register is initialized in the uniform superposition state given in (Alg.~1, L5):
\begin{equation}
|s\rangle = H^{\otimes n}|0^n\rangle = \frac{1}{\sqrt{N}}\sum_x |x\rangle, \qquad N=2^n.
\label{eq:s}
\end{equation}
The diffuser operator acts only on the $n$ variable qubits; see \Cref{eq:D}.
\begin{equation}
D = 2|s\rangle\langle s| - I = 
H^{\otimes n}\bigl(2|0^n\rangle\langle 0^n| - I\bigr)H^{\otimes n}.
\label{eq:D}
\end{equation}
One Grover step is $G = D U_F$.

Let $W = \{x : F(x) = 1\}$, the set of satisfying assignments, with $|W| = K$. Define $\lvert w \rangle$ as the uniform superposition over $W$, and let $\lvert w_\perp \rangle$ be its orthogonal complement.
A single Grover iteration $G = D U_F$ performs a planar rotation in
$\operatorname{span}\{\lvert w \rangle, \lvert w_\perp \rangle\}$ by angle $2\theta$, where $\sin^2 \theta = \frac{K}{N}$.

After $r$ iterations, the success probability of measuring a satisfying assignment is $\sin^2\!\big((2r+1)\theta\big)$.
When there are $K$ solutions, the near-optimal choice is expressed by \Cref{eq:r}.
\begin{equation}
r^\star \approx \left\lfloor \frac{\pi}{4\theta} \right\rfloor
= \Theta\!\left(\sqrt{\frac{N}{K}}\right).
\label{eq:r}
\end{equation}

\begin{figure}[t!]
\begin{algorithm}[H]
\caption{Q-LEAK BBHT witness search for a leakage-verification CNF}
\label{alg:grover-bbht-sca}
\begin{algorithmic}[1]
\Require CNF $F(x)$ encoding a side-channel verification case over $n$ Boolean variables
\Ensure SAT/UNSAT and satisfying assignment $x^\star$ (if found)
\Function{GroverSCA\_BBHT}{$F, n$}
  \State $N \gets 2^n$, $k \gets 1$, $\lambda \gets 8/7$
  \Repeat
    \State choose integer $r$ uniformly at random from $\{0,\dots,k-1\}$
    \State prepare $n$-qubit search register in $|s\rangle = H^{\otimes n}\lvert 0^n\rangle$
    \For{$i = 1$ to $r$}
      \State apply oracle $U_F \gets \Call{BuildOracle}{F}$ on search+ancilla qubits
      \State apply diffusion operator $D$ on the search register
    \EndFor
    \State measure search register to obtain $x \in \{0,1\}^n$
    \If{$F(x) = 1$} \Comment{classical CNF evaluation}
      \State \Return SAT, $x^\star \gets x$
    \EndIf
    \State $k \gets \min(\lambda k, \sqrt{N})$
  \Until{$k > \sqrt{N}$ or resource budget exhausted}
  \State \Return UNSAT, $\bot$
\EndFunction

\Function{BuildOracle}{$F$}
  \State let $F=\bigwedge_{j=1}^{m}C_j$, $C_j=(\ell_{j1}\lor\cdots\lor\ell_{jp_j})$
  \State allocate clause-satisfaction bits $c_1,\dots,c_m$ and flag $f$ in $\lvert0\rangle$
  \For{each clause $C_j$}
    \State compute $c_j\gets C_j$ \Comment{$c_j=1$ iff $C_j$ is satisfied}
  \EndFor
  \State compute $f \gets \bigwedge_{j=1}^{m}c_j$
  \Comment{$f=1$ iff $F(x)=1$}
  \State apply a phase flip controlled on $f$
  \Comment{$\lvert x\rangle \mapsto (-1)^{F(x)}\lvert x\rangle$}
  \State uncompute $f$ and all $c_j$ to reset the ancillas to $\lvert 0\rangle$
  \State \Return $U_F$
\EndFunction
\end{algorithmic}
\end{algorithm}
\end{figure}

In SAT settings, the value of $K$ is unknown. To address this, we adopt the Boyer–Brassard–Høyer–Tapp (BBHT) strategy~\cite{boyer1998tight} to randomize the number of Grover iterations. We initialize $k\leftarrow 1$ and choose a growth factor $\lambda>1$ (e.g., $8/7$), as shown in \Cref{alg:grover-bbht-sca}-L2.  
We then repeat the following steps:
(i) Sample $r\sim\text{Unif}\{0,1,\ldots,k-1\}$ (L4),
(ii) Prepare $|\psi\rangle = (D U_F)^r |s\rangle$, applies $r$ Grover steps, calling \textsc{BuildOracle} once per step and then the diffuser (L6–9),
(iii) Measure to obtain $x$ (L10). A classical CNF check $F(x)$ decides success (L11–12).
If $F(x)=1$, the case is declared SAT (L12). 
If not found, we \textit{perform another try}, where we increase $k\leftarrow \min(\lambda k,\lceil\sqrt{N}\rceil)$ (L14) and repeat until a query budget is reached (L15).
The BBHT method achieves expected query complexity $O(\sqrt{N/K})$ for all $K \ge 1$.
If $K = 0$ (UNSAT), then $U_F$ reduces to the identity, no rotation toward marked states
occurs, and UNSAT is reported after the procedure completes.

To certify that a case is UNSAT based on the outputs of \textit{Q-LEAK}, all measurement outcomes across BBHT trials are aggregated into a single bitstring histogram and compared against the ideal uniform distribution $1/N$ using the standard Pearson $\chi^2$ goodness-of-fit test~\cite{rana2015chi}. If uniformity is not rejected and no bitstring exhibits a persistent spike across many Grover iterations, the data are consistent with the absence of marked states. 
Under the hypothesis that no marked states exist, the observed histogram has the expected uniform shape, and the risk that this is incorrect (marked states exist but are missed) is bounded by a small parameter $\delta$; since \textit{Q-LEAK} is a probabilistic sampler, this statistical layer turns a visually flat histogram into an explicit, tunable bound on the probability of a false UNSAT claim.

For a quantitative guarantee, a confidence level $1-\delta$ is fixed, and Hoeffding's inequality~\cite{hoeffding1963probability} is applied to the empirical frequencies over all $S$ shots, yielding a deviation radius $\varepsilon = \sqrt{\ln(1/\delta)/(2S)}$ such that, with probability at least $1-\delta$, each true probability lies within $\varepsilon$ of its observed value. If the maximum observed probability satisfies $p_{\max} \le 1/N + \varepsilon$, then, with the same confidence, the total probability mass on all satisfying assignments is bounded by $K/N \le \varepsilon$,
where $K$ is the number of satisfying states and $N$ is the size of the search space. In this regime, the instance is reported as UNSAT at confidence level $1-\delta$.

Compared to classical CDCL solvers, which explore assignments via branching and propagation, \textit{Q-LEAK} amplifies the amplitudes of all possible assignment states in superposition. Through repeated oracle and diffusion operations, the state vector is rotated toward the satisfying subspace, thereby transforming the combinatorial search into a sequence of oracle-and-diffuser iterations.





\section{Experimental Setup}


The experimental evaluation focuses on simple, fully controlled CNF cases that fit within the current simulation and hardware limits of Q-LEAK. These are generated leakage-checking templates rather than a public industrial benchmark suite. All benchmarks are derived from a common power side-channel verification template, instantiated with different semantic constraints on the leakage relation and on the keys. Concretely, we use a single-bit state ($n_{bits}=1$), one unrolled time step ($T=1$), and the BIT leakage model. This setting is intentionally minimal: it makes the complete quantum distribution easy to inspect, permits exact comparison against a classical SAT solver, and exposes the oracle and noise costs before scaling to larger ciphers. It does not yet demonstrate full higher-order masking, multi-cycle unrolling, or HW/HD leakage. Those requested experiments require additional implementation and hardware resources and are therefore framed as future work.

Within this minimal setting, we distinguish three representative cases. Case 1 enforces that the leakage of the two executions differ (property mode \textit{``notequal''}), without imposing any additional constraint on the keys and with a fixed plaintext bit of one. This configuration yields a quantum circuit with a width of $17$ qubits and a depth of $40$. Case 2 keeps the same structural parameters but instead enforces equal leakage between the two runs (property mode \textit{``equal''}), again without requiring the keys to differ. The resulting quantum circuit has a width of $21$ qubits and a depth of $44$. Case 3 combines the \textit{``notequal''} leakage condition with an explicit key-difference constraint, such that any satisfying assignment must exhibit both a key change and a leakage change. In this case, the quantum circuit has a width of $27$ qubits and a depth of $46$. These three cases provide a small yet meaningful set of SAT/UNSAT behavior and oracle structures, which is used to study the functionality of the Q-LEAK solver because we vary only the logical content of the constraints, while keeping the scale of the cases controlled. As shown in Table~\ref{tab:leakage-instances}, each case is characterized by $(n,m,K)$, where $n$ is the variable count, $m$ is the clause count, and $K$ represents how many assignments violate the leakage condition, along with the corresponding circuit width and depth. 

As a classical baseline, we solve the exact same CNFs with Kissat~\cite{biere2024cadical}. On these small instances, Kissat is expected to be effectively instantaneous; therefore, the comparison is a correctness and reproducibility check rather than a runtime win claim for Q-LEAK at the current implementation. The baseline verifies every reported quantum witness and confirms the UNSAT control. Larger CDCL-hard leakage instances, projected resource estimates on such instances, and hybrid decompositions are important future work.

\begin{table}[t!]
\centering
\caption{Representative leakage-checking cases and corresponding quantum circuit sizes.}
\label{tab:leakage-instances}
\resizebox{\columnwidth}{!}{%
\begin{tabular}{lcccccc}
\toprule
\textbf{Case} & \textbf{Mode / Constraint} & \textbf{$n$} & \textbf{$m$} & \textbf{$K$} & \textbf{\makecell{Qubits\\ (width)}} & \textbf{\makecell{Depth /\\ Grover step}} \\
\midrule
Case 1 (Fig.~\ref{SAT_UNSAT}a \& \ref{IBM_Hardware}a) &
\makecell{notequal leakage,\\ keys unconstrained, $p{=}1$} &
5 & 11 & 2 & 17 & 40 \\
\hline
Case 2 (Fig.~\ref{SAT_UNSAT}b \& \ref{IBM_Hardware}b) &
\makecell{equal leakage,\\ keys unconstrained, $p{=}1$} &
6 & 14 & 2 & 21 & 44 \\
\hline
Case 3 (Fig.~\ref{SAT_UNSAT}c) &
\makecell{notequal leakage,\\ keys differ, $p{=}1$} &
7 & 19 & 2 & 27 & 46 \\
\hline
UNSAT (Fig.~\ref{SAT_UNSAT}d) &
no marked items ($K{=}0$) &
5 & 9 & 0 & 15 & 32 \\
\bottomrule
\end{tabular}}
\end{table}

The quantum experiments are implemented using the PennyLane framework~\cite{bergholm2018pennylane}. For simulation, all runs are performed on an NVIDIA GeForce RTX~4090 GPU with $24\,564$ MiB of memory. We rely on PennyLane’s \texttt{lightning.gpu }device as the statevector backend. Each circuit evaluation uses $2\,000$ shots to estimate measurement outcomes. For every benchmark case, the entire procedure is repeated $10$ times, and the results are averaged to mitigate the statistical fluctuations arising from measurement noise and randomized choice of BBHT iteration counts.

We also evaluate \textit{Q-LEAK} on the 156-qubit \texttt{ibm\_marrakesh} device accessed via the IBM Quantum cloud service. For each case, we executed multiple runs with $2\,000$ 
shots per attempt. The entire procedure was repeated five times, and the reported results correspond to averages over these five repetitions. We deliberately report the hardware runs without aggressive error mitigation so that the observed false-positive peaks reveal the impact of near-term noisy devices.

\section{Results and Discussion}
\label{sec:resultsdiscussion}

\begin{figure*}[ht!]
    \centering
    \includegraphics[width=\linewidth]{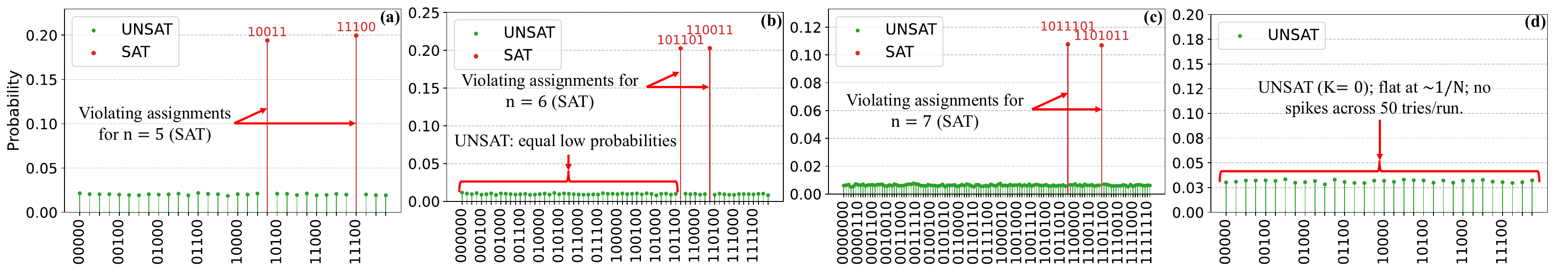}
    \caption{Q-LEAK results on power side-channel CNF benchmarks. Four CNF cases are plotted (a) number of variables $n=5$, two solutions; clear spikes. (b)~$n=6$, two solutions; higher peaks due to better r hit. (c) $n=7$, two solutions; peaks are lower as $N$ grows with fixed shots. (d) UNSAT control; flat near $1/N$, no spikes, consistent with no satisfying assignment. Results are averaged over 10 runs.}
    \label{SAT_UNSAT}
\end{figure*}





This section analyzes the three benchmark cases under increasing search spaces $N = 2^n$ with a fixed number of satisfying assignments $K=2$. Circuit sizes are reported for a single Grover iteration ($r=1$): the quantum oracles require $17/21/27$ qubits and depths $40/44/46$ for $n=5/6/7$ qubits, respectively. Kissat solves the same CNFs and provides the reference witness set. Because the tested CNFs are intentionally tiny, the relevant comparison is agreement of the returned assignments and resource accounting, not measured speedup over classical SAT. Fig.~\ref{SAT_UNSAT} illustrates the measured probability over bitstrings, averaged over 10 runs: green dots correspond to non-satisfying assignments, while red spikes indicate SAT assignments.

Fig.~\ref{SAT_UNSAT}a corresponds to Case~1 ($n=5$, $m=11$, $N=32$). We observe two clear violating assignments, each with probability around $0.20$, while all other bitstrings form a low background. Under BBHT, Q-LEAK identifies a witness in $1.2$ tries on average. The assignments returned by Q-LEAK coincide with the Kissat outputs, namely $[1,1,1,0,0]$ and $[1,0,0,1,1]$.

Fig.~\ref{SAT_UNSAT}b shows Case~2 ($n=6$, $m=14$, $N=64$), where the two classically verified witnesses produce visible probability spikes. The algorithm requires approximately two tries on average, although some runs require the BBHT parameter $k$ to grow before a favorable value of $r$ is sampled. The quantum witnesses match the Kissat results, $[101101]$ and $[110011]$.

Fig.~\ref{SAT_UNSAT}c presents the results of Case 3 ($n=7$, $m=19$, \mbox{$N=128$}). With the larger search space, BBHT frequently samples small values of $r$ (often $r{=}1$), which produce visible but lower peaks, with probabilities around $0.11$ and quick successes. Occasional trials with $r{=}4$ further amplify these peaks. The average number of tries is $1.3$. As in the previous cases, the SAT assignments returned by \textit{Q-LEAK} are consistent with the classical solver outputs, namely $[1011101]$ and $[1101011]$.

A geometric view clarifies these observations. Let $\sin^2\theta = K/N$. Each Grover iteration rotates the state in the $(|w\rangle, |w_\perp\rangle)$ plane; after $r$ steps, the success probability is $\sin^2((2r{+}1)\theta)$. As $N$ doubles with fixed $K$, $\theta$ decreases and the optimal number of iterations $r^\ast \approx \frac{\pi}{4\theta} = \Theta(\sqrt{N/K})$ grows (approximately $1,2,3,4$ for the three SAT cases). Since $K$ is unknown, BBHT samples $r$ uniformly from $\{0,\ldots,k{-}1\}$ while gradually increasing $k$ ($k\leftarrow \min(\lambda k,\sqrt{N})$). This ensures that a good $r$ is reached in a small number of trials. This behavior is consistent with our experimental results, where more $k$-growth is required at $N{=}64$, and at $N{=}128$ larger values of $r$ occasionally yield more pronounced peaks.

All $N$ candidate assignments are initialized in superposition. The oracle is realized as a compute-phase-uncompute circuit, where the variable register is entangled with clause ancillas, all clause bits and their AND are computed, a global phase flip is applied if and only if all clauses are satisfied, and the computation is then uncomputed to disentangle the ancillas. This selective phase flip induces quantum interference in the subsequent diffuser: amplitudes associated with non-solutions interfere destructively, while amplitudes corresponding to satisfying assignments interfere constructively, producing the red spikes. Formally, $U_F$ multiplies satisfying basis states by $-1$. The composition of $U_F$ with the diffuser $D$ implements the pair of reflections that yield the Grover rotation described above.


The number of oracle calls follows the expected $\Theta(\sqrt{N/K})$ scaling (realized through BBHT's randomized choice of $r$). The cost of each oracle call is determined by its depth and width, which grow with CNF structure (number of literals and clauses, as well as embedded subcircuits for XOR and population count in HW/HD encodings). In the BIT model with $T{=}1$ regime, the qubit count and circuit depth grow gently with $n$, and \textit{Q-LEAK} consistently finds a witness within $1$-$4$ tries across all runs, in line with the theoretical behavior for small $K$ and moderate $N$. However, larger HW/HD models and multi-cycle unrolling would introduce additional XOR, popcount, and transition-count subcircuits. These costs are precisely the bottleneck that must be reduced before practical quantum advantage on noisy hardware can be expected.

Fig.~\ref{SAT_UNSAT}d shows an UNSAT instance with $n = 5$ variables, $m = 9$ clauses, and $K = 0$ satisfying assignments. The compiled oracle circuit uses $15$ qubits and has depth $32$ for a single Grover step. We collect $S = 10 \times 2000$ shots over $10$ runs, covering the $N = 32$ computational basis states. A $\chi^{2}$ goodness-of-fit test against the uniform distribution $1/N$ yields $\chi^{2} = 30.81$ with $31$ degrees of freedom, which is consistent with the expected no-marked-state distribution. Using Hoeffding's inequality at confidence $1-\delta = 0.99$ gives $\varepsilon \approx 1.1 \times 10^{-2}$; the maximum observed frequency satisfies $p_{\max} \le 1/N + \varepsilon$, which in our experiment corresponds to $0.033 \le 0.042$. Hence, the total marked mass is bounded by $K/N \le \varepsilon$. We therefore report the quantum data as statistical evidence consistent with UNSAT at $99\%$ confidence.

\begin{figure}[t!]
    \centering
    \includegraphics[width=\linewidth]{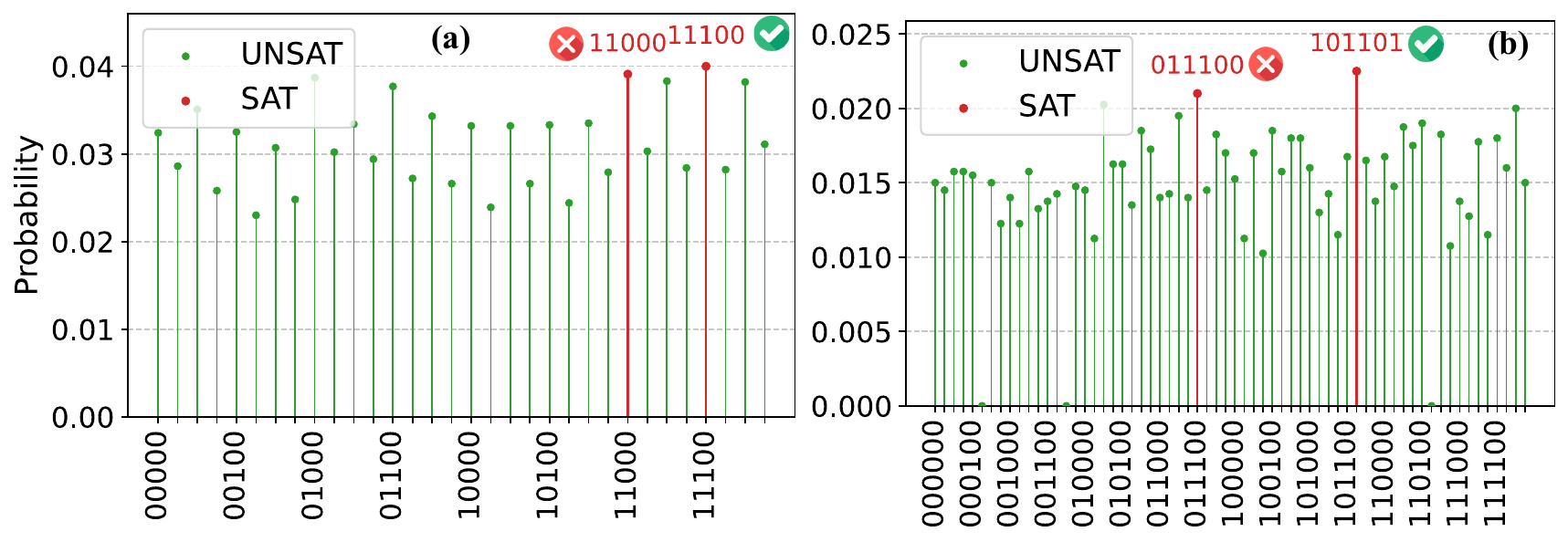}
    \caption{Hardware evaluation of Q-LEAK on \texttt{ibm\_marrakesh} (156 qubits). Measured bitstring probabilities for (a) case 1 and (b) case 2. Red stems mark candidate SAT assignments; \checkmark\ indicates classically verified solutions (11100, 101101), \texttimes\ marks spurious peaks (11000, 011100) attributed to noise. Green stems show the UNSAT-like background. Results are averaged over 5 runs.}
    \label{IBM_Hardware}
\end{figure}

\textbf{Evaluation on Quantum Hardware: } Fig.~\ref{IBM_Hardware} presents the hardware measurement histograms for Case 1 (Fig.~\ref{IBM_Hardware}a) and case 2 (Fig.~\ref{IBM_Hardware}b). In both cases, one correct SAT assignment appears as the highest red bar (\checkmark: $11100$ in Fig.~\ref{IBM_Hardware}a; $101101$ in Fig.~\ref{IBM_Hardware}b). Additional tall bars also appear (\texttimes: 11000, 011100), which are false positives caused by hardware noise, routing, and readout errors.
Compared with the noise-free simulation, the hardware spectra are flatter, and the gap between the true SAT peak and the UNSAT-like green background is reduced. As a result, BBHT requires more growth steps and retries to reach the correct peak. Nevertheless, each hardware run still produces at least one verified SAT bitstring per case, confirming that the end-to-end \textit{Q-LEAK} pipeline functions correctly on real quantum hardware.

\vspace{-4pt}
\section{Projected Hard-Instance and Circuit-Resource Analysis}
\label{sec:projected-resources}
The Kissat comparison in Sec.~\ref{sec:resultsdiscussion} is intentionally a correctness baseline: the tested CNFs are too small to make any classical solver struggle. This section therefore addresses the scalability question without claiming unavailable large-scale quantum experiments. We analyze a hypothetical hard residual SAT family that remains after classical simplification, propagation, and problem-specific decomposition. Let $n_{\mathrm{eff}}$ denote the number of residual Boolean variables that still require search and let $m$ denote the remaining clauses. The raw design size may be much larger than $n_{\mathrm{eff}}$; in a hybrid flow, CDCL, structural hashing, cone-of-influence reduction, or manual decomposition should first remove easy structure, and Q-LEAK is meaningful only on the high-entropy residual.

For a residual with $K\ge1$ leakage witnesses, the BBHT/Grover query count is
\vspace{-4pt}
\begin{equation}
R_Q(n_{\mathrm{eff}},K)\approx \frac{\pi}{4}\sqrt{\frac{2^{n_{\mathrm{eff}}}}{K}}.
\label{eq:projected-rq}
\end{equation}
For the current oracle style, the logical width is
\vspace{-4pt}
\begin{equation}
q_{\log}=n_{\mathrm{eff}}+m+1,
\label{eq:qlog}
\end{equation}
where the terms are search variables, one clause ancilla per CNF clause, and one CNF flag. For a 3-CNF-like residual, using a clean-ancilla decomposition, a conservative serial Toffoli-class count per Grover step is
\vspace{-4pt}
\begin{equation}
C_Q(n_{\mathrm{eff}},m)\approx 4m+2(m-1)+(2n_{\mathrm{eff}}-3)=6m+2n_{\mathrm{eff}}-5.
\label{eq:cq}
\end{equation}
The three terms correspond to clause compute/uncompute, global AND compute/uncompute, and the diffuser multi-control. This estimate is logical and pre-layout: it excludes native-gate decomposition constants, SWAP routing, magic-state factories, error correction, measurement latency, and repeated BBHT trials. It is therefore not a wall-clock estimate; it is a transparent resource model for the part of the computation that Grover accelerates.

\begin{table}[t!]
\centering
\caption{Projected logical resources for hard residual 3-CNF instances using the current one-ancilla-per-clause Q-LEAK oracle, with conservative $K=1$ and clause density $m=8n_{\mathrm{eff}}$. $R_Q$ is the near-optimal Grover iteration count, $C_Q$ is the Toffoli-class count per Grover step from eq.\ref{eq:cq}, and $T_Q$ is the projected quantum logical operation count. We compare this against a residual classical solver model $T_C(n_{\mathrm{eff}})$, where $\alpha$ is the measured or projected effective classical exponent. The threshold $\alpha_{\min}$ is the classical exponent that would match the quantum logical operation count.}
\label{tab:projected-resources}
\begingroup

\resizebox{0.5\textwidth}{!}{%
\begin{tabular}{rrrrrrr}
\toprule
$n_{\mathrm{eff}}$ & $m$ & $q_{\log}$ & $R_Q$ & $C_Q$ & $R_QC_Q$ & $\alpha_{\min}$ \\
\midrule
16 & 128 & 145 & $ 2.01\times 10^{2} $ & 795 & $ 1.60\times 10^{5} $ & $1.08$ \\
32 & 256 & 289 & $ 5.15\times 10^{4} $ & 1595 & $ 8.21\times 10^{7} $ & $0.82$ \\
64 & 512 & 577 & $ 3.37\times 10^{9} $ & 3195 & $ 1.08\times 10^{13} $ & $0.68$ \\
80 & 640 & 721 & $ 8.64\times 10^{11} $ & 3995 & $ 3.45\times 10^{15} $ & $0.65$ \\
128 & 1024 & 1153 & $ 1.45\times 10^{19} $ & 6395 & $ 9.27\times 10^{22} $ & $0.60$ \\
\bottomrule
\end{tabular}%
}
\endgroup
\vspace{-8pt}
\end{table}

\begin{figure}[t!]
\centering
\begingroup
\setlength{\fboxsep}{0pt}
\includegraphics[width=0.5\textwidth]{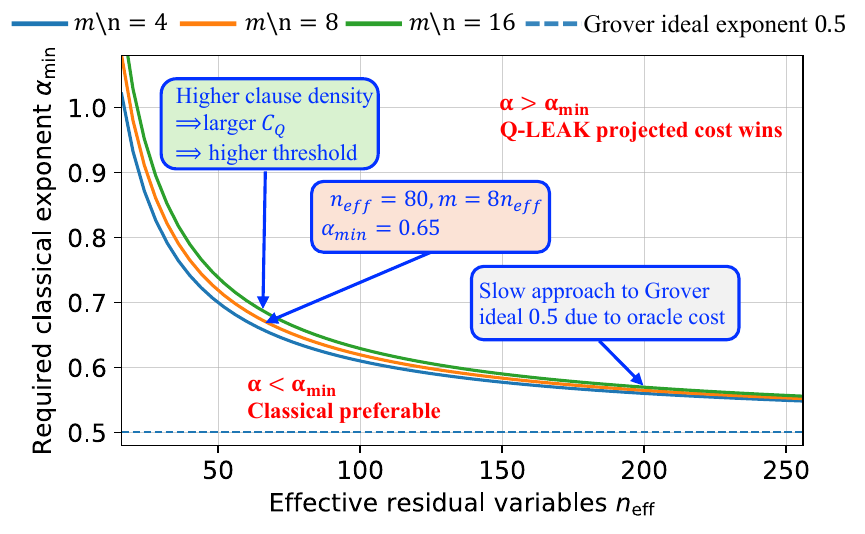}
\endgroup
\caption{Projected hard-instance threshold for Q-LEAK oracle-weighted advantage. Each curve gives the break-even classical exponent $\alpha_{\min}$. Above a curve $(\alpha>\alpha_{\min})$, Q-LEAK has lower projected logical cost; below it $(\alpha<\alpha_{\min})$, the classical or hybrid solver remains preferable. The annotated $n_{\mathrm{eff}}=80$, $m=8n_{\mathrm{eff}}$ example gives $\alpha_{\min}\approx0.65$. Higher clause density increases $C_Q$ and therefore raises the threshold, while the curves approach the ideal Grover exponent $0.5$ slowly due to oracle overhead.}
\label{fig:hard-instance-projection}
\vspace{-8pt}
\end{figure}

Table~\ref{tab:projected-resources} and Fig.~\ref{fig:hard-instance-projection} show the practical meaning of the claimed advantage. Let $T_Q=R_QC_Q$ denote the projected quantum logical operation count for the residual instance, and let $T_C(n_{\mathrm{eff}})\approx2^{\alpha n_{\mathrm{eff}}}$ denote the empirical or projected cost of a classical CDCL/hybrid solver on the same residual. The value $\alpha_{\min}=\log_2(T_Q)/n_{\mathrm{eff}}$ is therefore the break-even classical exponent obtained from $T_C=T_Q$. If the measured classical exponent satisfies $\alpha>\alpha_{\min}$, then the classical residual search is projected to exceed the Q-LEAK logical count; if $\alpha<\alpha_{\min}$, the classical or hybrid classical method remains preferable.

At $n_{\mathrm{eff}}=80$ and $m=640$, the model requires $721$ logical qubits and about $8.6\times10^{11}$ Grover steps, i.e., about $3.5\times10^{15}$ Toffoli-class logical operations before fault-tolerant overhead. This is far beyond our current simulators and the NISQ hardware used in Sec.~\ref{sec:resultsdiscussion}, so we do not present it as an executable experiment. However, the search exponent is still reduced from $80$ to $40$ in the oracle-query term. A hard classical residual whose effective scaling exponent is above $\alpha_{\min}\approx0.65$ would be larger than the quantum logical count in this model; if CDCL learning or decomposition lowers the residual exponent below that threshold, the classical or hybrid classical method remains preferable.

The separation among the curves in Fig.~\ref{fig:hard-instance-projection} reflects oracle overhead. Since $C_Q(n_{\mathrm{eff}},m)=6m+2n_{\mathrm{eff}}-5$, increasing the clause density $m/n_{\mathrm{eff}}$ directly increases the cost of each Grover step. Therefore denser residual CNFs require a larger classical exponent $\alpha$ before Q-LEAK becomes favorable. Conversely, as $n_{\mathrm{eff}}$ grows, the dominant Grover term approaches the ideal $2^{n_{\mathrm{eff}}/2}$ scaling, so $\alpha_{\min}$ gradually approaches $0.5$, but only slowly because the oracle factor $C_Q$ remains present in $T_Q=R_QC_Q$.

This analysis also clarifies how hard classical instances should be used in future evaluations. The relevant comparison is not raw Kissat time on the tiny quantum-executable cases, but a paired study on larger leakage CNFs: first measure or project the residual classical exponent after preprocessing, then compute $q_{\log}$, $C_Q$, and $R_QC_Q$ for the same residual. In a hybrid decomposition, classical reasoning should reduce the formula to independent hard blocks and Q-LEAK should be applied only where the residual entropy is high enough to overcome oracle overhead. Thus the advantage claimed in this paper is conditional but concrete: Q-LEAK changes the exponent of a hard residual counterexample search, while the resource table exposes the logical qubit and Toffoli costs that must be made fault-tolerant before practical speedup can be realized.

\section{Limitations and Future Work}
Our evaluation focuses on one-bit BIT leakage, $T=1$ unrolling, and controlled templates with $n \leq 7$ search variables. These experiments validate the complete workflow, including the CNF-to-oracle construction, BBHT integration, classical post-checking, and hardware execution path. The present results address functional correctness and end-to-end implementation rather than large-scale performance. Scalability to full masked AES, higher-order masking, multi-cycle traces, and HW/HD leakage, as well as runtime crossover against classical SAT solvers, remains to be established through larger and more complex benchmarks.

The next experimental steps are: (i) connect Q-LEAK to measurement-calibrated leakage predicates derived from power or EM traces; (ii) generate multi-cycle BIT/HW/HD CNFs from realistic masked datapaths; (iii) compare Kissat/CryptoMiniSat runtimes and learned-clause behavior against projected quantum oracle resources on those same CNFs; and (iv) evaluate mitigation strategies such as readout correction, transpilation-aware qubit mapping, and lower-depth multi-control decompositions.

The main resource bottleneck is the oracle. With the current construction, the logical width is roughly $n+m+1$ qubits before routing and fault-tolerance overhead, and the depth grows with the number of clauses, literal controls, and any XOR/popcount circuitry introduced by the leakage model. Consequently, a useful implementation for large verification cases would likely require many more logical qubits than the small demonstrations here, and substantially more physical qubits after error correction. A precise ``how many qubits'' answer depends on the target cipher, the unrolling bound $T$, the leakage model, the clause encoding, and the error-correction code. A prospective avenue for future research is the development of an oracle compression framework together with a hybrid classical–quantum decomposition scheme.

\section{Conclusion}


Formal verification of large leakage-analysis circuits remains computationally challenging. We proposed \textit{Q-LEAK}, a leakage-aware Grover/BBHT SAT framework that encodes the countermeasure check as a CNF oracle, applies amplitude amplification over the variable qubits, and uses BBHT to handle the unknown number of solutions. We implemented it in PennyLane with the \texttt{lightning.gpu} backend and on the \texttt{ibm\_marrakesh} device. The present study is a proof of concept: the benchmarks are small, controlled BIT-leakage templates designed to validate the end-to-end mapping and resource behavior, rather than immediate speedup over classical SAT solvers on noisy hardware. Across $n=\{5,6,7\}$ with $K$ violating assignments, \textit{Q-LEAK} recovered valid witnesses consistent with the classical SAT baseline within $1$--$4$ tries in most runs; the UNSAT control produced near-flat spectra with no marked-state amplification.

The results show that side-channel verification constraints can be compiled into a quantum oracle and probed using amplitude amplification, yielding clear SAT peaks when solutions exist and uniform outcomes otherwise. The main limitations stem from the rising cost of oracle construction as the number of clauses increases, which pushed our computational resources to their limit at about 30 qubits simulations, along with BBHT overhead and noise sensitivity for real hardware. Our implementation validates correctness on benchmarks and, on real IBM hardware, recovers at least one correct assignment per case despite noise, albeit with lower contrast and more attempts than in simulation.
This work opens a path toward faster verification of power side-channel countermeasures and marks a step toward practical security validation of larger circuits. Future work will focus on leaner clause encodings, richer leakage models, hybrid pruning with classical SAT, and hardware experiments with error mitigation.


\section*{Acknowledgments}
This work was supported in part by the NYUAD Center for Cyber Security (CCS), funded by Tamkeen under the NYUAD Research Institute Award G1104.

\bibliographystyle{IEEEtran}
\bibliography{refs.bib}

@inproceedings{grover1996fast,
  title={A fast quantum mechanical algorithm for database search},
  author={Grover, Lov K},
  booktitle={Proceedings of the twenty-eighth annual ACM symposium on Theory of computing},
  pages={212--219},
  year={1996}
}

@article{bennett1997strengths,
  title={Strengths and weaknesses of quantum computing},
  author={Bennett, Charles H and Bernstein, Ethan and Brassard, Gilles and Vazirani, Umesh},
  journal={SIAM journal on Computing},
  volume={26},
  number={5},
  pages={1510--1523},
  year={1997},
  publisher={SIAM}
}

@incollection{mai2011side,
  title={Side channel attacks and countermeasures},
  author={Mai, Ken},
  booktitle={Introduction to hardware security and trust},
  pages={175--194},
  year={2011},
  publisher={Springer}
}

@inproceedings{kocher1999differential,
  title={Differential power analysis},
  author={Kocher, Paul and Jaffe, Joshua and Jun, Benjamin},
  booktitle={Annual international cryptology conference},
  pages={388--397},
  year={1999},
  organization={Springer}
}

@inproceedings{brier2004correlation,
  title={Correlation power analysis with a leakage model},
  author={Brier, Eric and Clavier, Christophe and Olivier, Francis},
  booktitle={International workshop on cryptographic hardware and embedded systems},
  pages={16--29},
  year={2004},
  organization={Springer}
}

@book{mangard2007power,
  title={Power analysis attacks: Revealing the secrets of smart cards},
  author={Mangard, Stefan and Oswald, Elisabeth and Popp, Thomas},
  year={2007},
  publisher={Springer}
}

@article{messerges2002examining,
  title={Examining smart-card security under the threat of power analysis attacks},
  author={Messerges, Thomas S and Dabbish, Ezzat A and Sloan, Robert H},
  journal={IEEE transactions on computers},
  volume={51},
  number={5},
  pages={541--552},
  year={2002},
  publisher={IEEE}
}

@inproceedings{nikova2006threshold,
  title={Threshold implementations against side-channel attacks and glitches},
  author={Nikova, Svetla and Rechberger, Christian and Rijmen, Vincent},
  booktitle={International conference on information and communications security},
  pages={529--545},
  year={2006},
  organization={Springer}
}

@inproceedings{chari1999towards,
  title={Towards sound approaches to counteract power-analysis attacks},
  author={Chari, Suresh and Jutla, Charanjit S and Rao, Josyula R and Rohatgi, Pankaj},
  booktitle={Annual International Cryptology Conference},
  pages={398--412},
  year={1999},
  organization={Springer}
}

@inproceedings{barthe2015verified,
  title={Verified proofs of higher-order masking},
  author={Barthe, Gilles and Bela{\"\i}d, Sonia and Dupressoir, Fran{\c{c}}ois and Fouque, Pierre-Alain and Gr{\'e}goire, Benjamin and Strub, Pierre-Yves},
  booktitle={Annual International Conference on the Theory and Applications of Cryptographic Techniques},
  pages={457--485},
  year={2015},
  organization={Springer}
}

@inproceedings{bloem2018formal,
  title={Formal verification of masked hardware implementations in the presence of glitches},
  author={Bloem, Roderick and Gro{\ss}, Hannes and Iusupov, Rinat and K{\"o}nighofer, Bettina and Mangard, Stefan and Winter, Johannes},
  booktitle={Annual International Conference on the Theory and Applications of Cryptographic Techniques},
  pages={321--353},
  year={2018},
  organization={Springer}
}

@inproceedings{hadzic2021cocoalma,
  title={COCOALMA: A versatile masking verifier},
  author={Hadzic, Vedad and Bloem, Roderick},
  booktitle={21st International Conference on Formal Methods in Computer-Aided Design: FMCAD 2021},
  pages={14--23},
  year={2021}
}

@inproceedings{schreiber2024trusted,
  title={Trusted scalable SAT solving with on-the-fly LRAT checking},
  author={Schreiber, Dominik},
  booktitle={27th International Conference on Theory and Applications of Satisfiability Testing (SAT 2024)},
  pages={25--1},
  year={2024},
  organization={Schloss Dagstuhl--Leibniz-Zentrum f{\"u}r Informatik}
}

@article{rana2015chi,
  title={Chi-square test and its application in hypothesis testing},
  author={Rana, Rakesh and Singhal, Richa},
  journal={Journal of the practice of cardiovascular sciences},
  volume={1},
  number={1},
  pages={69--71},
  year={2015},
  publisher={Medknow}
}

@article{hoeffding1963probability,
  title={Probability inequalities for sums of bounded random variables},
  author={Hoeffding, Wassily},
  journal={Journal of the American statistical association},
  volume={58},
  number={301},
  pages={13--30},
  year={1963},
  publisher={Taylor \& Francis}
}

@article{boyer1998tight,
  title={Tight bounds on quantum searching},
  author={Boyer, Michel and Brassard, Gilles and H{\o}yer, Peter and Tapp, Alain},
  journal={Fortschritte der Physik: Progress of Physics},
  volume={46},
  number={4-5},
  pages={493--505},
  year={1998},
  publisher={Wiley Online Library}
}

@article{yang2023efficient,
  title={Efficient quantum circuit synthesis for SAT-oracle with limited ancillary qubit},
  author={Yang, Shuai and Zi, Wei and Wu, Bujiao and Guo, Cheng and Zhang, Jialin and Sun, Xiaoming},
  journal={IEEE Transactions on Computer-Aided Design of Integrated Circuits and Systems},
  volume={43},
  number={3},
  pages={868--877},
  year={2023},
  publisher={IEEE}
}

@article{vinod2021finding,
  title={Finding solutions to the integer case constraint satisfiability problem using Grover’s algorithm},
  author={Vinod, Gayathree M and Shaji, Anil},
  journal={IEEE Transactions on Quantum Engineering},
  volume={2},
  pages={1--13},
  year={2021},
  publisher={IEEE}
}

@article{lin2023parallel,
  title={A parallel and distributed quantum SAT solver based on entanglement and quantum teleportation},
  author={Lin, Shang-Wei and Wang, Tzu-Fan and Chen, Yean-Ru and Hou, Zhe and San{\'a}n, David and Teo, Yon Shin},
  journal={arXiv preprint arXiv:2308.03344},
  year={2023}
}

@article{eldib2014formal,
  title={Formal verification of software countermeasures against side-channel attacks},
  author={Eldib, Hassan and Wang, Chao and Schaumont, Patrick},
  journal={ACM Transactions on Software Engineering and Methodology (TOSEM)},
  volume={24},
  number={2},
  pages={1--24},
  year={2014},
  publisher={ACM New York, NY, USA}
}

@article{een2005minisat,
  title={Minisat-a SAT solver with conflict-clause minimization},
  author={Een, Niklas},
  journal={Proc. Theory and Applications of Satisfiability Testing (SAT 05)},
  year={2005}
}

@article{leventi2021cryptominisat,
  title={CryptoMiniSat switches-optimization for solving cryptographic instances},
  author={Leventi-Peetz, A-M and Zendel, Oliver and Lennartz, Werner and Weber, Kai},
  journal={arXiv preprint arXiv:2112.11484},
  year={2021}
}

@inproceedings{eldib2014smt,
  title={SMT-based verification of software countermeasures against side-channel attacks},
  author={Eldib, Hassan and Wang, Chao and Schaumont, Patrick},
  booktitle={International Conference on Tools and Algorithms for the Construction and Analysis of Systems},
  pages={62--77},
  year={2014},
  organization={Springer}
}

@inproceedings{agrawal2002side,
  title={The EM side—channel (s)},
  author={Agrawal, Dakshi and Archambeault, Bruce and Rao, Josyula R and Rohatgi, Pankaj},
  booktitle={International workshop on cryptographic hardware and embedded systems},
  pages={29--45},
  year={2002},
  organization={Springer}
}

@article{lou2021survey,
  title={A survey of microarchitectural side-channel vulnerabilities, attacks, and defenses in cryptography},
  author={Lou, Xiaoxuan and Zhang, Tianwei and Jiang, Jun and Zhang, Yinqian},
  journal={ACM Computing Surveys (CSUR)},
  volume={54},
  number={6},
  pages={1--37},
  year={2021},
  publisher={ACM New York, NY, USA}
}

@article{spreitzer2017systematic,
  title={Systematic classification of side-channel attacks: A case study for mobile devices},
  author={Spreitzer, Raphael and Moonsamy, Veelasha and Korak, Thomas and Mangard, Stefan},
  journal={IEEE communications surveys \& tutorials},
  volume={20},
  number={1},
  pages={465--488},
  year={2017},
  publisher={IEEE}
}

@article{lyu2018survey,
  title={A survey of side-channel attacks on caches and countermeasures},
  author={Lyu, Yangdi and Mishra, Prabhat},
  journal={Journal of Hardware and Systems Security},
  volume={2},
  number={1},
  pages={33--50},
  year={2018},
  publisher={Springer}
}

@article{bergholm2018pennylane,
  title={Pennylane: Automatic differentiation of hybrid quantum-classical computations},
  author={Bergholm, Ville and Izaac, Josh and Schuld, Maria and Gogolin, Christian and Ahmed, Shahnawaz and Ajith, Vishnu and Alam, M Sohaib and Alonso-Linaje, Guillermo and AkashNarayanan, B and Asadi, Ali and others},
  journal={arXiv preprint arXiv:1811.04968},
  year={2018}
}

@article{biere2024cadical,
  title={CaDiCaL, Gimsatul, IsaSAT and Kissat entering the SAT competition 2024},
  author={Biere, Armin and Faller, Tobias and Fazekas, Katalin and Fleury, Mathias and Froleyks, Nils and Pollitt, Florian},
  journal={Proc. of SAT Competition},
  pages={8--10},
  year={2024}
}

@inproceedings{coron2018formal,
  title={Formal verification of side-channel countermeasures via elementary circuit transformations},
  author={Coron, Jean-S{\'e}bastien},
  booktitle={International Conference on Applied Cryptography and Network Security},
  pages={65--82},
  year={2018},
  organization={Springer}
}
\end{document}